\documentclass[sigplan,10pt,nonacm]{acmart}

\setcopyright{none}
\renewcommand\footnotetextcopyrightpermission[1]{}
\acmConference{}{}{}
\acmBooktitle{}

\newcommand{\sys}{\textsc{ReFlux}\xspace}
\newenvironment{smitemize}{\begin{list}{$\bullet$}{\setlength{\parsep}{0pt}\setlength{\topsep}{0pt}\setlength{\leftmargin}{2pc}\setlength{\itemsep}{1pt}}}
  {\end{list}}

\usepackage{courier}
\usepackage{xspace}
\usepackage{booktabs}
\usepackage{microtype}
\usepackage{enumitem}
\usepackage{url}
\hypersetup{pdfauthor={Yanpeng Hu; Yiwei Yang; Yusheng Zheng; Estabon Ramos; Jianchang Su; Andi Quinn; Wei Zhang}}

\begin{document}

\title{\sys: Reversible Compute Placement for CXL-Enabled Storage}

\author{Yanpeng Hu}
\authornote{Yanpeng Hu and Yiwei Yang contributed equally to this work.}
\affiliation{\institution{ShanghaiTech University}
  \country{China}}
\email{huyp@shanghaitech.edu.cn}

\author{Yiwei Yang}
\authornotemark[1]
\affiliation{\institution{UC Santa Cruz}
  \country{USA}}
\email{yyang363@ucsc.edu}

\author{Yusheng Zheng}
\affiliation{\institution{UC Santa Cruz}
  \country{USA}}
\email{yzhen165@ucsc.edu}

\author{Estabon Ramos}
\affiliation{\institution{UC Santa Cruz}
  \country{USA}}
\email{esiramos@ucsc.edu}

\author{Jianchang Su}
\affiliation{\institution{University of Connecticut}
  \country{USA}}
\email{jianchang.su@uconn.edu}

\author{Andi Quinn}
\affiliation{\institution{UC Santa Cruz}
  \country{USA}}
\email{aquinn1@ucsc.edu}

\author{Wei Zhang}
\affiliation{\institution{University of Connecticut}
  \country{USA}}
\email{wei.13.zhang@uconn.edu}
\begin{abstract}
Static offload to computational storage devices is brittle:
device-side processors throttle under sustained thermal load,
and opaque, vendor-specific interfaces inflate adoption cost to
the point where no computational storage platform has achieved
broad deployment.
We argue that storage-side compute should be \emph{reversible}:
individual pipeline stages should migrate between host and
device at runtime, beneath standard interfaces that require
zero application modification.
We present \sys, which realizes this principle on CXL SSDs
by decomposing I/O-path logic into migratable \emph{storage
actors} compiled to WebAssembly.
Actors share state through coherent CXL.mem regions so that
only lightweight control state ($\sim$8\,KB) moves during
migration; a thermal-aware scheduler triggers per-stage
drain-and-switch when device temperature or queue pressure
rises, uploading compute-intensive actors while leaving
I/O-bound stages on the device.
Our evaluation on an FPGA-based CXL SSD prototype and two
production CSDs shows that \sys sustains 2$\times$ higher
throughput than thermally throttled CSDs under 30-minute
sustained writes, delivers 3--4$\times$ inference throughput
under KV-cache pressure, and reduces host CPU utilization
by 65\% through MWAIT-based notification.
\end{abstract}
\maketitle
\pagestyle{plain}

\section{Introduction}
\label{sec:intro}

Data movement between storage and compute remains a persistent
bottleneck in modern systems.
Traditional data systems spend substantial I/O-path time on
compression, logging, and integrity checks~\cite{barbalace2021computational},
while emerging LLM systems increasingly rely on storage-backed
offloading of model state during training and of KV caches during
long-context inference~\cite{jang2024smartinfinity,pan2025instattention}.
Near-storage execution is therefore attractive: instead of
copying data across PCIe for every transformation, the system can
run storage-path logic close to the media.
When these transformations remain on the host, the system pays
both the computation cost and the data-movement cost needed to
reach it.
Computational storage devices (CSDs) pursue this idea by placing
processing capability next to flash~\cite{snia2019computational,barbalace2021computational}.

Broad deployment of CSDs has nevertheless stalled for two
structural reasons.
First, device-side compute runs inside a tight SSD thermal and
power envelope; under sustained load, static offload turns into a
throughput cliff with no recovery path.
Our measurements on production CSDs show that this failure mode
is routine under sustained writes rather than an isolated stress
artifact.
Second, commercial devices expose vendor-specific toolchains and
heterogeneous controller ISAs, forcing per-device ports and
breaking code portability.
The result is a brittle model: offload helps when conditions are
favorable, but once device temperature rises or platforms change,
applications inherit the complexity.

Prior systems address only parts of this problem.
Static offload systems and products~\cite{jun2015bluedbm,ruan2019insider,amd2024smartssd}
keep computation on the device and cannot reverse placement when
thermal conditions deteriorate.
Coarse thermal-management schemes~\cite{song2023coco} can reduce
device activity, but they still do not make placement reversible
at the granularity of individual storage-path stages.
Host--device co-execution systems such as
$\lambda$-IO~\cite{yang2023lambdaio} choose between host and
device at request granularity, but cannot recall or split
in-flight work and rely on explicit coherence management.
CXL-based storage systems~\cite{directcxl2022,kwon2025blockbyte}
show that coherent, byte-addressable device memory reduces access
latency, but they focus on \emph{data} placement rather than
\emph{compute} placement.
None of these systems provides runtime-reversible, per-stage
compute placement beneath standard storage interfaces.

Our key observation is that storage-side compute should be
\emph{reversible}: the system should move storage-path logic
between host and device as runtime conditions change, without
exposing that choice to applications.
Realizing this idea requires migrating execution quickly enough
to remain on the storage fast path, executing the same logic on
both sides without per-platform rewrites, and deciding what to
move without introducing thrashing.
We call this principle \emph{reversible compute placement}.

We present \sys, which realizes reversible compute placement on
CXL-enabled SSDs.
\sys keeps shared actor state in coherent CXL.mem-backed PMR and
migrates only lightweight control state between host and device.
CXL coherence makes this practical: migration cost tracks actor
control state rather than shared data state.
It expresses I/O-path logic as \emph{storage actors} compiled to
WebAssembly, so the same actor binary runs on both sides without
application-visible changes.
A thermal-aware scheduler applies drain-and-switch migration at
per-stage granularity, uploading only the compute-intensive stage
when device temperature or queue pressure rises while leaving
I/O-bound stages in place.
All mechanisms sit beneath standard POSIX and block interfaces,
so existing applications run unchanged.

Our contributions are:
\begin{description}[leftmargin=0pt, labelwidth=2em, labelsep=0.3em, itemsep=1pt, parsep=0pt, topsep=2pt]
\item \textbf{Reversible compute placement.}
  We introduce a computational-storage model in which
  storage-path logic moves between host and device at runtime to
  bypass the thermal cliffs of static offload while preserving
  application transparency.
\item \textbf{Transparent, zero-copy migration.}
  We co-design a WebAssembly actor runtime with coherent CXL
  memory so that actors migrate by moving only lightweight
  control state while shared state remains stationary in PMR.
\item \textbf{Per-stage thermal-aware scheduling.}
  We design a drain-and-switch scheduler that migrates
  individual pipeline stages rather than whole requests, allowing
  the system to upload only thermally intensive actors under
  sustained load.
\end{description}
We implement \sys on an FPGA-based CXL~2.0 SSD prototype and
evaluate it with unmodified applications.
\sys turns hard thermal cliffs into elastic trade-offs,
sustaining 2$\times$ higher throughput over thermally throttled
CSDs, delivering 3--4$\times$ inference throughput under
sustained KV-cache pressure, and reducing host CPU utilization
by 65\% at low queue depths through MWAIT-based notification.

\section{Background}
\label{sec:bg}

\subsection{Computational Storage and the Portability Wall}
\label{sec:bg_csd}

Computational storage devices (CSDs) embed processing elements
alongside flash media so that data-intensive operations---compression,
encryption, filtering---execute near the data rather than
traversing the host I/O
stack~\cite{acharya1998active,riedel2001active}.
Modern incarnations differ in both execution substrate and
programming model: Samsung SmartSSD~\cite{amd2024smartssd}
couples an FPGA with enterprise NAND via the Vitis toolchain,
ScaleFlux CSD1000~\cite{scalefluxcsd2024} uses a fixed-function
ASIC for transparent compression, and NGD
Newport~\cite{newport2019} exposes embedded ARM cores
programmable through a custom Linux userspace.
The ecosystem is correspondingly fragmented.
Each platform exposes a vendor-specific interface with little code
portability~\cite{snia2019computational}; controllers are
diverging across ARM and RISC-V
(e.g., Samsung's BM9K1~\cite{samsung2026bm9k1}); and
conventional PCIe still isolates device memory from the host, so
host--device data sharing requires explicit DMA and
software-managed cache coherence.

\subsection{WebAssembly as a Systems Substrate}
\label{sec:bg_wasm}

WebAssembly (WASM) was originally designed as a portable
compilation target for
browsers~\cite{haas2017wasm}; its sandboxing guarantees and
compact bytecode have since attracted systems-level
adoption~\cite{bytecodealliance2023wasmtime,yang2025mvvmdeployaiagentssecurely}.
For heterogeneous storage runtimes, WASM is relevant because it
offers three properties in one execution format.
\emph{Portability:} a WASM module compiles once and executes on
any platform with an AOT-capable runtime, eliminating the need
for per-device rebuilds when host and device run different
processors.
\emph{Lightweight instantiation:} module startup completes in
tens of microseconds, small enough to sit on the storage fast
path~\cite{jangda2019not}.
\emph{Compact, serializable state:} a WASM instance's control
state---call stack, local variables, linear memory
base---is isolated and small (typically a few kilobytes),
making it amenable to checkpoint and migration at runtime.

\subsection{CXL.mem and Coherent Device Memory}
\label{sec:bg_cxl}

Compute Express Link (CXL) multiplexes three sub-protocols
over a PCIe physical layer~\cite{cxl-spec}.
Of the three, CXL.mem is most relevant to this work: it lets
the host issue load/store operations to device-attached memory
with hardware-managed coherence, eliminating the explicit flushes
and invalidations required by conventional PCIe
DMA~\cite{sun2023demystifying}.
A CXL SSD combines commodity NAND and an NVMe controller with
a CXL controller that exposes a coherent Persistent Memory
Region (PMR) to the host via
CXL.mem~\cite{cxlssd2023,samsung2022memoryssd}.
The host maps PMR into its physical address space;
device-embedded cores operate over the same region.
Because coherence is maintained in hardware, data placed in PMR
is immediately visible to both sides without software
intervention---a property that enables shared state to remain
stationary when computation migrates between host and device.

\subsection{Limitations of Static Offload and Co-execution}
\label{sec:bg_coexec}

Existing computational-storage architectures are either fully
static or limited to one-shot dispatch.
Static offload (SmartSSD, ScaleFlux) pins computation to the
device; when thermal throttling activates, throughput collapses
with no recourse.
Filesystem-level coordination schemes for compression-based
SSDs~\cite{song2023coco} can reduce device-side compression
work, but they still make coarse placement decisions and cannot
migrate arbitrary storage-path stages at runtime.
$\lambda$-IO~\cite{yang2023lambdaio} is the closest prior
work in host--device co-execution: it profiles
execution time and dispatches each \emph{entire request} to the
faster side via a unified
\texttt{pread\_\allowbreak{}lambda}/\texttt{pwrite\_\allowbreak{}lambda}
interface.
However, $\lambda$-IO's dispatch model has two structural
constraints.
First, the scheduling unit is a whole request: once dispatched,
every pipeline stage (decompression, checksum, index lookup)
runs on the same side, even if only one stage is the thermal
bottleneck.
Second, the dispatch decision is irrevocable---a request in
flight cannot be recalled or split mid-stream.
Consistency between host and device relies on explicit
flush/invalidate operations, precluding zero-copy shared state.
These constraints point to an unmet need: a computational-storage
architecture that supports \emph{per-stage}, \emph{runtime-reversible}
compute placement with coherent shared state.

\section{Motivation}
\label{sec:motv}

To design a computational-storage architecture that sustains
performance in production, we must first understand why existing
platforms fail under sustained load and heterogeneous deployment
constraints.
This section empirically characterizes representative production
CSDs and derives the design requirements behind \sys.

\subsection{The Fallacy of Static Offload}
\label{sec:fallacy}

\begin{figure}[t]
\centering
\includegraphics[width=\columnwidth]{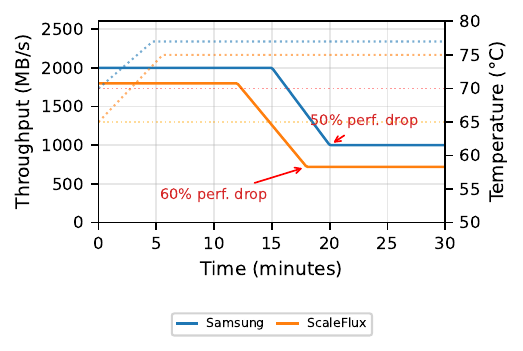}
\caption{Thermal throttling under sustained writes on two production CSDs.}
\label{fig:thermal}
\end{figure}

The dominant CSD model relies on \emph{static offload}~\cite{amd2024smartssd,scalefluxcsd2024}:
it pins storage-path computation to the device-side processors and
treats the SSD as the permanent execution site for that logic.
While this model can accelerate favorable workloads, it becomes
fundamentally brittle under sustained I/O because the device must
absorb both flash activity and offloaded compute within a strict
thermal envelope.
Enterprise SSDs typically operate within a tight thermal
budget~\cite{kioxia2023edsff}, and the OCP datacenter NVMe
specification mandates throttling above
77$^\circ$C~\cite{ocp2023nvme} to protect NAND reliability.
CSDs increase the thermal pressure on that envelope by colocating
flash activity with additional FPGA or embedded compute on the
device.

To quantify how this thermal wall manifests in practice, we
subject two production CSDs to sustained sequential writes, a
pattern routinely generated by datacenter tasks such as LSM
compaction, checkpoint, and backup.
Figure~\ref{fig:thermal} shows the effect on these platforms.
SmartSSD exhibits multi-stage throttling: the NVMe controller
throttles at 70$^\circ$C with 50\% throughput loss, the FPGA
reduces frequency at 93$^\circ$C, and clock gating activates at
97$^\circ$C~\cite{amd2024smartssd}.
ScaleFlux throttles at 65$^\circ$C with 60\% degradation.
In multi-tenant deployments, concurrent workloads sharing the
same device accelerate thermal buildup further, compounding
the throttling problem.
Any viable abstraction therefore needs an escape path that can
move storage-path computation back to the host before hardware
throttling becomes the only remaining control mechanism.

\emph{Observation~1:} Under sustained load, device-side compute
becomes a liability, not an optimization; once pinned to the
device, there is no escape path.

\subsection{The Adoption Barrier of New Storage Hardware}
\label{sec:tco}

Beyond runtime brittleness, computational storage faces an
adoption barrier rooted in interface design.
Samsung SmartSSD requires the Vitis/SDAccel toolchain; ScaleFlux
exposes only fixed-function compression; NGD Newport demands a
custom embedded-Linux workflow~\cite{snia2019computational}.
Code written for one device cannot run on another, and the
ecosystem fragments further with Samsung's newly announced
BM9K1---the first client SSD with a RISC-V controller---which
departs from the ARM-based designs of the previous
generation~\cite{samsung2026bm9k1}.
Without cache coherency, shared data between host and device
also requires explicit synchronization, negating much of the
near-data benefit~\cite{barbalace2021computational}.
Persistent memory illustrates the risk: Optane's dual-mode
programming model demanded months of application refactoring,
and most deployments defaulted to Memory Mode that sacrificed
persistence~\cite{intel2022optane,vmware2020pmem}.
Interface opacity was among the factors that limited
adoption and contributed to Optane's
discontinuation~\cite{intel2022optane}.
The common root cause is \emph{interface opacity}: each new
storage technology asked applications to adopt proprietary
libraries or programming models.
\sys takes the opposite approach---all benefits are realized
beneath standard POSIX and block interfaces, with zero
application modification (\S\ref{sec:design}).

The ISA heterogeneity between host (x86) and device (ARM in
current products, RISC-V in emerging ones) compounds the problem:
any execution abstraction used for reversible compute must allow
the same storage logic to run on both sides without requiring
developers to maintain separate codebases.
Reversible compute placement therefore demands a unified
execution substrate that meets three requirements
simultaneously: the ability to run unmodified on heterogeneous
processors, microsecond-scale instantiation on the storage fast
path, and compact, serializable execution state that can be
checkpointed and migrated at runtime.

\subsection{The CXL Opportunity}
\label{sec:cxl_opportunity}

\begin{figure}[t]
\centering
\includegraphics[width=\columnwidth]{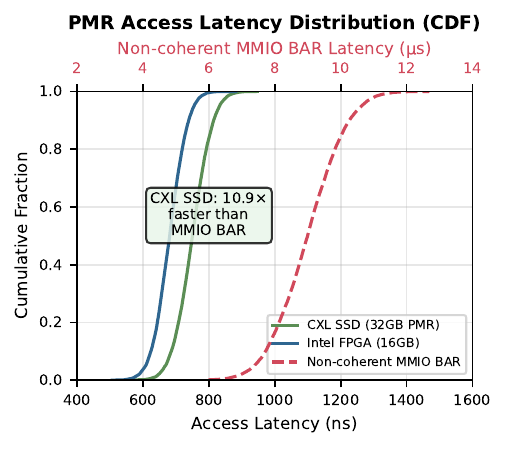}
\caption{PMR access latency distribution (CDF) compared with non-coherent MMIO BAR.}
\label{fig:pmr_latency}
\end{figure}

The CXL protocol (\S\ref{sec:bg_cxl}) provides the coherent
shared memory that prior substrates lacked.
The CXL SSD's 32\,GB PMR delivers 750\,ns median
latency---10.9$\times$ better than non-coherent MMIO BAR
access (Figure~\ref{fig:pmr_latency})---and 22\,GB/s sequential
throughput; once the working set exceeds the 32\,GB PMR capacity,
latency falls to NVMe levels, confirming that PMR must be treated
as a managed hot tier.
For fine-grained 8\,B writes through the filesystem stack,
CXL.mem achieves 5.4\,$\mu$s versus 38\,$\mu$s (SmartSSD) and
80.6\,$\mu$s (ScaleFlux), where the latency is dominated by the
read-modify-write overhead of buffered I/O.
Raw \texttt{mmap} writes bypass the filesystem entirely:
8\,B and 512\,B writes complete in 0.47--0.61\,$\mu$s, compared
with 18.39\,$\mu$s for buffered writes and 53.78\,$\mu$s for
\texttt{O\_DIRECT}---confirming that cache-coherent load/store
eliminates sector-granularity overhead.

However, CXL alone is not sufficient.
Naive polling consumes 100\% host CPU at low queue depths, and
PMR capacity is limited.
These measurements show that CXL provides a practical coherent
shared-memory substrate for fine-grained storage-path execution,
but they do not by themselves determine where computation should
run or how execution should move between host and device.
Coherent shared memory is the \emph{enabler} for reversible
compute, not the solution itself.

\emph{Observation~2:} CXL coherency makes reversible compute
practical by providing low-latency shared memory, but the system
still needs a sandboxed runtime, a migration protocol, and
migration-aware scheduling to realize it.

\section{\sys System Architecture}
\label{sec:design}

\begin{figure*}[t]
\centering
\includegraphics[width=0.85\textwidth]{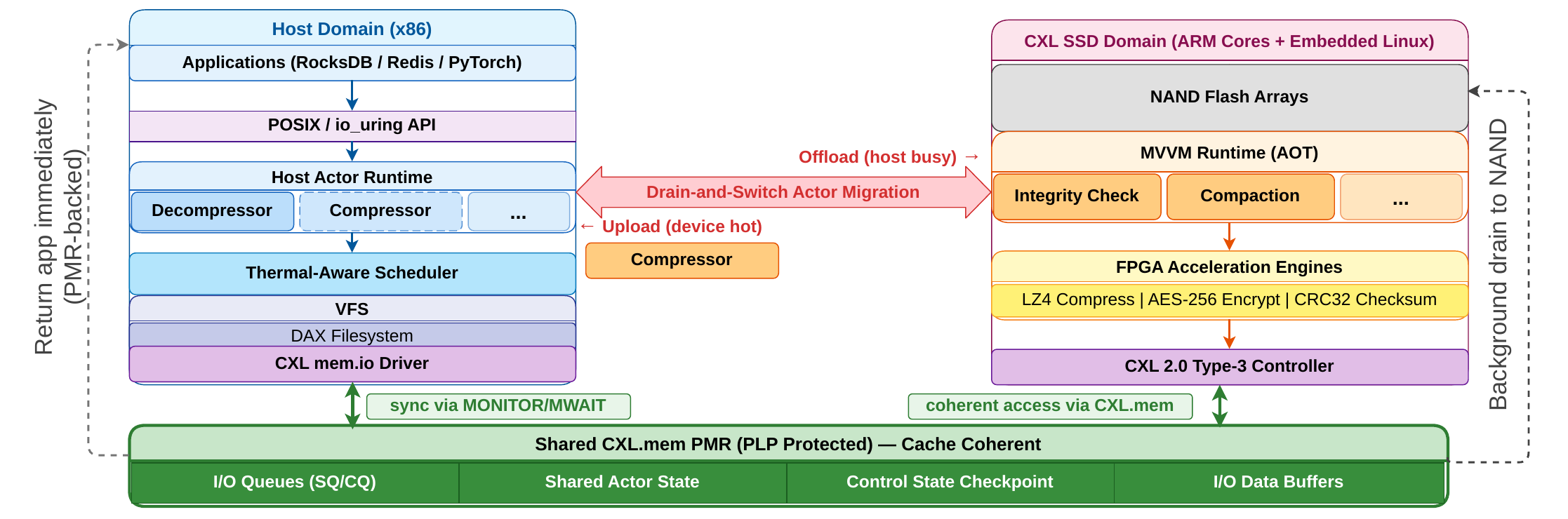}
\caption{\sys architecture overview; the compression actor is being uploaded from the device to the host due to rising device temperature.}
\label{fig:cxl_architecture}
\end{figure*}

The observations in \S\ref{sec:motv} establish both the need for
reversible placement and the constraint under which it must be
implemented.
Observation~1 shows that static offload fails unless the system
can upload computation before thermal throttling collapses
throughput, while Observation~2 shows that CXL makes such upload
practical only if migration cost scales with control state rather
than shared state.
These observations point to three challenges that a reversible
computational-storage system must address.
First, the thermal cliffs of static offload demand that
computation be \emph{uploadable} on the storage fast path without
stalling the I/O pipeline (Challenge~1).
Second, because host and device may run different ISAs (x86 on
the host, ARM on the device in our prototype), the execution
substrate must allow the same storage logic to run on both sides
\emph{without requiring developers to maintain per-device code};
today's CSD ecosystem forces separate toolchains for each
platform~\cite{snia2019computational}, and \sys must avoid
imposing similar burden on users (Challenge~2).
Third, the system must decide \emph{which} actor to migrate and
\emph{when}, balancing device thermal headroom against host
utilization without thrashing (Challenge~3).
The central design consequence is that shared state must remain
stationary while execution moves; otherwise migration would copy
state-sized data and erase the latency advantage that makes
upload practical in the first place.
\sys addresses all three with four design goals:
\begin{smitemize}
\item \emph{Application transparency.}
  Existing applications continue to use standard POSIX and block
  interfaces without any code modification; all benefits from CXL
  SSDs are realized beneath those interfaces.
\item \emph{Microsecond-scale migration.}
  Actor migration must complete fast enough ($<$50\,$\mu$s) to sit
  on the storage fast path without stalling I/O.
\item \emph{No-thrashing scheduling.}
  The migration policy must improve performance, not degrade it;
  migration overhead must not exceed the throughput loss it prevents.
\item \emph{Crash safety.}
  Migration must not lose application data or leave actors in an
  inconsistent state after power failure.
\end{smitemize}

\subsection{Transparent Application Interface}

Figure~\ref{fig:cxl_architecture} illustrates the two-domain
architecture.
Host applications issue I/O through POSIX or
\texttt{io\_uring};
\sys interposes between \texttt{io\_uring} and the page cache,
attaching storage actors to each request's processing path.
The device domain runs ARM cores with a separate Linux instance
and a dual-JIT WebAssembly runtime.
A central 32\,GB PMR, backed by power-loss-protection (PLP)
capacitors, serves as coherent shared memory;
I/O queues, data buffers, and shared actor state reside here.
An FPGA acceleration layer provides engines for compression (LZ4),
encryption (AES-256), checksumming (CRC32), and other
data-transformation primitives.
NAND arrays provide bulk capacity with asynchronous background
draining from PMR to flash.
As the figure illustrates, when device temperature approaches
the throttling threshold, a compute-intensive actor such as
the compressor is uploaded to the host while I/O-bound actors
(integrity check, compaction) remain on the device.

The key to application transparency is placing the I/O control
path in coherent PMR.
\sys modifies the \texttt{io\_uring} fast
path~\cite{axboe2019iouring} to attach per-request metadata
describing the actor pipeline and references into shared state.
The \texttt{io\_uring} rings map naturally onto PMR-resident
queues---both use shared-memory rings with producer/consumer
pointers and support batched, zero-copy semantics.
Each submission queue entry carries a 32-byte descriptor
identifying the actors to run, their input/output buffers in PMR,
and a handle to a per-request state blob; a 4-bit opcode selects
a predefined pipeline (compress, encrypt, checksum, passthrough)
and a flags word enables optional stages.
Page-cache hooks allocate selected pages from the PMR DAX node
and tag each cache page with an actor identifier, placement
hints, and an epoch counter.
The epoch counter is essential for correctness: if a host thread
reads a page whose epoch has advanced, it detects an in-flight
migration and retries after the migration completes.
Traditional \texttt{read}/\texttt{write} and \texttt{mmap} paths
also work via VFS hooks, but without the MWAIT power optimization
available on the \texttt{io\_uring} fast path.

\subsection{Storage Actors and WebAssembly Execution}
\label{sec:design_wasm}

\sys expresses I/O-path logic as \emph{storage
actors}~\cite{agha1986actors}: small modules bound to fixed
positions in a per-request dataflow pipeline.
Each actor consumes pages or records from its predecessor,
consults and updates shared state, and forwards results
downstream.
This dataflow restriction simplifies migration (an actor's
interface is determined by its pipeline position) and lets the
scheduler reason about per-stage cost without tracking arbitrary
message graphs.
Typical actors include decompressors, integrity checkers,
encryptors, JSON parsers, and predicate evaluators; a read of
compressed, checksummed log segments might invoke integrity check
$\rightarrow$ decompress $\rightarrow$ decode, while background
compaction invokes key rewriting and index construction.

Each actor has two kinds of state.
\emph{Control state}---instruction pointer, call stack, local
variables---is small, private, and serializable.
\emph{Shared state}---counters, per-range metadata, LRU lists,
statistics---lives in the CXL-mapped PMR so both host and device
actors see the same view.
During migration, only control state moves; shared state stays in
place and ownership transfers via a small metadata protocol that
guarantees single-writer semantics.

Challenge~2 requires the same actor binary to run on both host
and device without recompilation or per-platform toolchains.
\sys addresses this through a WebAssembly-based execution
substrate paired with a Double-JIT translation pipeline that
bridges the ISA gap at the device boundary.
Storage actors are compiled once to WASM modules implementing a
restricted ABI for page access, metadata lookup, and logging.
On the host, modules run inside in-kernel WASM sandboxes or
user-space helper processes, with ahead-of-time (AOT) compilation
to x86 keeping overhead close to native.
On the device, the same WASM modules execute inside a dual-JIT
runtime built around a three-stage translation
pipeline: the frontend parses incoming x86 actor binaries
into WebAssembly's linear memory model, the middleend emits
validated WASM bytecode via a Wasmtime-based backend with a
WASIX-compatible libc, and the backend JIT-compiles WASM to the
device's native ISA (ARM in our prototype) with a persistent code
cache.
Because the backend targets WASM rather than a specific device
ISA, the same pipeline can in principle extend to future
controller architectures such as RISC-V without
application-visible changes.
The runtime isolates actor instances through per-instance
linear memory regions with PMR access mediated by a narrow
syscall-like interface, which lets \sys run untrusted actors
while preserving device integrity.

Three alternatives to WebAssembly merit discussion.
\emph{Native plugins} offer the best raw performance but require
a separate build per ISA and provide no memory isolation without
hardware support.
\emph{eBPF}~\cite{mccanne1993bpf} is attractive for in-kernel
storage extensions---XRP~\cite{zhong2022xrp} reports
2.5$\times$ speedups for index lookups---but the eBPF verifier
imposes strict constraints (bounded loops, 512-byte stack, no
dynamic heap, 33-invocation tail-call chains) that rule out
multi-stage actor pipelines with variable-sized state (compression
dictionaries, LRU metadata).
\emph{Lightweight VMs}~\cite{agache2020firecracker} provide strong
isolation but incur millisecond-scale startup and require a guest
kernel, impractical for per-request instantiation.
WebAssembly~\cite{haas2017wasm} occupies the practical middle
ground: one actor format for both host and device, memory-safe sandboxing
without kernel privilege, and compact control state that is easy
to checkpoint during drain-and-switch.
Section~\ref{sec:eval_wasm} quantifies the overhead of this
abstraction across I/O-dominated and compute-intensive workloads.
Device-side actors use AOT compilation to avoid JIT warmup,
keeping the runtime footprint under 32\,MB.

\subsection{Zero-Copy Drain-and-Switch Migration}
\label{sec:design_migration}

Migration is the mechanism that makes upload practical on the
storage fast path.
When the scheduler decides to move an actor---whether due to
rising device temperature or host CPU headroom---the runtime executes a drain-and-switch protocol
that completes without dropping or replaying any request.

The protocol begins by redirecting new incoming requests to the
destination side (host or device), so that the source sees no
further arrivals.
The source then drains its in-flight requests to completion; because
actors are dataflow stages with bounded pipeline depth, this drain
phase is short---typically a few outstanding requests.
Once the source is quiescent, it serializes the actor's control
state (instruction pointer, call stack, local variables---roughly
8\,KB) into a designated region of the PMR and sets a \emph{ready}
flag.

The destination detects this flag through a MONITOR/MWAIT
instruction on the corresponding PMR cache line.
Because CXL.mem provides hardware-managed coherence, the device's
write to the ready flag propagates to the host's cache without
explicit flushes, and the monitoring core wakes with
sub-microsecond latency.
The same MWAIT mechanism underpins \sys's normal I/O completion
path: completion and submission queues are single-producer,
single-consumer rings in PMR, cache-line aligned and mapped
write-back; after submitting a request, the host sets up
monitoring on the expected completion entry and enters MWAIT with
a bounded timeout.
When the device completes the actor pipeline, its coherent write
to the completion entry propagates through CXL.cache/CXL.mem and
wakes the core.
At low queue depths, MWAIT drops CPU utilization from 100\% to
35\% with comparable P99 latency; at high queue depths,
traditional polling provides higher throughput, so the runtime
transitions from MWAIT to polling as inter-completion gaps shrink.

The destination then reconstructs the actor in a fresh WASM
sandbox, reattaches the shared state already resident in PMR, and
resumes execution.
The entire sequence completes in under 50\,$\mu$s on our prototype.
Two properties make this possible.
First, shared state never moves: counters, metadata, LRU lists,
and statistics remain in PMR and are immediately visible to the
destination through coherent load/store, so the migration cost is
proportional to the small control state, not the large shared
state.
Second, the MWAIT-based synchronization avoids both the high
latency of interrupts ($\sim$200\,$\mu$s) and the CPU waste of
busy-polling, making it practical to migrate actors at the
granularity of individual scheduling epochs (10\,ms) without
dedicating a core to migration coordination.
The 50\,$\mu$s migration cost is negligible relative to the
10\,ms scheduling epoch: each migration consumes $<$0.5\% of
the epoch budget.

\subsection{Thermal-Aware Per-Stage Scheduling}
\label{sec:design_sched}

Because drain-and-switch migration is cheap ($<$50\,$\mu$s), a
threshold-based scheme suffices to flatten thermal cliffs.
Every 10\,ms the scheduler samples host-side metrics (per-core
frequency, RAPL power draw, \texttt{io\_uring} queue depth) and
device-side metrics (controller temperature, core utilization,
queue depth), and for each actor~$a$ computes a placement score
$S(a) = \alpha \cdot T_\mathit{dev} + \beta \cdot U_\mathit{host}
       + \gamma \cdot Q_\mathit{dev}(a)$,
where $T_\mathit{dev}$ is normalized device temperature,
$U_\mathit{host}$ is host CPU utilization, and
$Q_\mathit{dev}(a)$ is $a$'s device-side queue depth.
We set $\alpha{=}0.5$, $\beta{=}0.3$, $\gamma{=}0.2$ through
offline grid search minimizing P99 latency; varying each weight
by $\pm$0.1 shifts P99 latency by less than 5\%.
When $S(a)$ exceeds an upload threshold~$\theta$, the actor is
uploaded; when it drops below~$\theta_\mathit{low}$ and the
device has headroom, the actor is offloaded back.
If both host and device are near saturation, the scheduler
throttles the request rate rather than migrating.

A critical design choice is the \emph{granularity} of migration.
$\lambda$-IO~\cite{yang2023lambdaio} dispatches at whole-request
granularity: once a request is assigned to host or device, every
stage runs on that side, so under thermal pressure the only
option is to redirect entire requests---even if only one
compute-intensive stage (e.g., compression) is the bottleneck.
\sys operates at \emph{per-stage} granularity: within a single
request pipeline, each actor can be placed independently, so
\sys uploads only the thermally intensive stage while keeping
I/O-bound stages (e.g., checksum verification) on the device.
The scheduler classifies actors by pipeline position:
latency-sensitive stages (metadata lookups, index updates) stay on
the host unless it is throttling, while background stages
(compression, compaction, log reformatting) are the primary
offload and upload candidates.
To prevent thrashing, each actor must remain at its current
location for at least 100\,ms before reconsideration, and at most
one actor migrates per epoch.
The scheduling overhead is dominated by the 10\,ms sampling
interval; reading hardware performance counters and CXL.io
temperature registers takes $<$1\,$\mu$s per epoch.

Not all actors are candidates for migration.
The scheduler classifies each actor by its \emph{pipeline role},
which is fixed at actor registration time: each predefined
pipeline (\S\ref{sec:design_wasm}) specifies whether an actor's
workload profile is I/O-dominated (compression, checksum, log
formatting) or compute-dominated (dense numeric kernels).
I/O-dominated actors are compiled to WASM and participate in
reversible placement---\S\ref{sec:eval_wasm} shows their WASM
overhead is negligible or negative compared to native.
Actors registered as compute-intensive are pinned to native
execution on one side and excluded from migration, avoiding the
high WASM translation cost on ALU-bound code.
In our prototype, all storage-path actors fall into the
WASM-migratable category.

\subsection{Durability and Crash Consistency}
\label{sec:design_durability}

\sys adopts an asynchronous durability model: writes complete
once data reaches PMR, which is backed by power-loss-protection
(PLP) capacitors that flush contents to NAND on power
failure~\cite{nvme2024base,ocp2023nvme}.
This is the same persistence guarantee that production datacenter
SSDs provide for their internal write buffers.
Background threads drain data from PMR to NAND asynchronously;
strict ordering requires explicit Global Persistent Flush (GPF)
barriers.

A crash during migration could leave an actor partially
checkpointed.
\sys uses a two-phase commit: the source writes a complete
checkpoint to PMR and sets a \emph{ready} flag; only after the
destination reconstructs the actor does it set an \emph{active}
flag.
A crash before \emph{ready} leaves the source in control; a crash
between \emph{ready} and \emph{active} triggers rollback to the
source.
Because shared state resides in PLP-protected PMR, no application
data is lost---only the $\sim$8\,KB control state may need
re-checkpointing.

\section{Implementation}
\label{sec:impl}

\subsection{Hardware Prototype}

Our \sys prototype is built on a CXL SSD platform combining an
Intel Agilex FPGA~\cite{intelfpga} with Samsung enterprise SSDs,
integrating 32\,GB PMR and controller DRAM exposed through
CXL~2.0.
Two PCIe~$\times$8 cores terminate CXL/PCIe links and expose two
BAR regions: BAR0 maps up to 128\,TB of CXL.mem-visible VMEM as
the byte-addressable aperture, while BAR2 provides an 8\,GB
region for configuration registers, queue state, and doorbells.
A MEM2NVME bridge translates between CXL.mem load/store semantics
and NVMe queue operations, using ordered Avalon semantics toward
the SSD controller and out-of-order AXI MCDMA within the
accelerator fabric.
From the host's perspective, a VMEM access appears as a single
coherent CXL.mem load/store; internally it maps to a sequence of
PCIe transactions and NVMe queue operations.

\subsection{Software Integration}

The host runs Ubuntu~22.04 with a Linux~6.2 kernel patched
for CXL~2.0 support.
We extend the CXL.mem driver to export the PMR as a DAX node and
to cooperate with \sys's page-cache allocation policies, so that
the buffer cache can selectively place metadata and actor state in
PMR rather than in host DRAM.
A small in-kernel control plane applies the scheduler's policies,
sampling per-core frequency, C-state residency, and memory
bandwidth through performance counters, and reading device
temperature and throughput via CXL.io configuration registers.
A user-space daemon handles configuration, tracing, and offline
policy replay.

The device runs a minimal Linux kernel on its embedded ARM cores,
with the MVVM runtime and v86 Double-JIT pipeline
(\S\ref{sec:design_wasm}) integrated into the vendor firmware
stack.
The three-stage x86$\to$WASM$\to$native ARM translation runs
entirely in device memory with a persistent code cache; for
production deployments, actors are distributed as AOT-compiled
native modules, keeping the device-side runtime footprint under
32\,MB.

All components described in \S\ref{sec:design}---the actor
runtime, drain-and-switch migration, MWAIT-based synchronization,
and the thermal-aware scheduler---are implemented and evaluated
on this prototype.

\section{Evaluation}
\label{sec:eval}

\subsection{Methodology}

Our evaluation compares three \emph{software paradigms},
isolating the contribution of each layer of abstraction.
Paradigms~B and~C run on the same FPGA-based CXL SSD prototype;
Paradigm~A uses two production CSDs as static-offload baselines.

\textbf{Paradigm~A (Static offload).}
Computation is pinned to the device's embedded cores and cannot
migrate.
For thermal characterization we measure
Samsung SmartSSD~\cite{amd2024smartssd} and ScaleFlux
CSD1000~\cite{scalefluxcsd2024} as representative fixed-offload
devices.
For controlled paradigm comparisons we emulate static offload on
the CXL SSD by pinning actors to device cores and disabling
migration.

\textbf{Paradigm~B (Request-level co-execution).}
The system profiles host and device execution time and dispatches
each incoming request to the faster side, but cannot recall or
split a request once dispatched.
We implement this on the CXL SSD by simulating
$\lambda$-IO~\cite{yang2023lambdaio} with whole-request
dispatch and no mid-flight migration.

\textbf{Paradigm~C (\sys).}
Per-stage actor placement with drain-and-switch migration,
thermal-aware scheduling, and MWAIT-based synchronization.

\paragraph{Experimental environment.}
We evaluate \sys on three platforms that represent the main
commercial design points of computational storage: our FPGA-based
CXL SSD prototype, Samsung SmartSSD as a reconfigurable
FPGA-backed CSD, and ScaleFlux CSD1000 as a fixed-function ASIC
CSD~\cite{amd2024smartssd,scalefluxcsd2024}.
The CXL prototype combines an Intel Agilex endpoint with Samsung
enterprise SSDs, exposes a 32\,GB PLP-protected PMR, and uses
controller-resident memory as a CMB.
All experiments run on a dual-socket Intel Xeon Gold 5418Y
(Sapphire Rapids) host with 256\,GB DDR4 and native CXL
support~\cite{intel2023xeon5418y}.
We use FIO~\cite{fio2022} for microbenchmarks and keep
ambient temperature stable across runs, and reset caches and
device state between runs so that reported throughput, latency,
CPU usage, and device telemetry reflect steady-state behavior
rather than residual state from prior trials.

Tests run 5~minutes after 2-minute warmup, averaged over 5~runs
with caches flushed between iterations.
All writes target the PLP-protected PMR;
data is durable at write completion, not merely buffered in
volatile DRAM.
The page cache is bypassed for micro-benchmarks
(\texttt{O\_DIRECT} or DAX).

We first demonstrate system-level thermal resilience
(\S\ref{sec:eval_thermal}), then show these benefits translate to
unmodified applications (\S\ref{sec:eval_e2e}), and finally
deconstruct the substrate and abstraction costs that enable
reversibility (\S\ref{sec:eval_decon}).
Three evaluation questions guide the presentation:
\begin{description}[leftmargin=0pt, labelwidth=2.5em, labelsep=0.3em, itemsep=1pt, parsep=0pt, topsep=2pt]
\item[\textbf{EQ1}] Under thermal stress, does reversible placement
  (Paradigm~C) sustain throughput where static offload (A) and
  request-level dispatch (B) collapse?
  (\S\ref{sec:eval_thermal})
\item[\textbf{EQ2}] Do these benefits hold end-to-end in
  unmodified applications such as DeepSeek inference?
  (\S\ref{sec:eval_e2e})
\item[\textbf{EQ3}] What CXL datapath characteristics and WASM
  abstraction costs enable reversible placement, and where do
  they limit it?
  (\S\ref{sec:eval_decon})
\end{description}

\subsection{Thermal Resilience under Sustained Load (EQ1)}
\label{sec:eval_thermal}

\begin{figure*}[t]
\centering
\includegraphics[width=\textwidth]{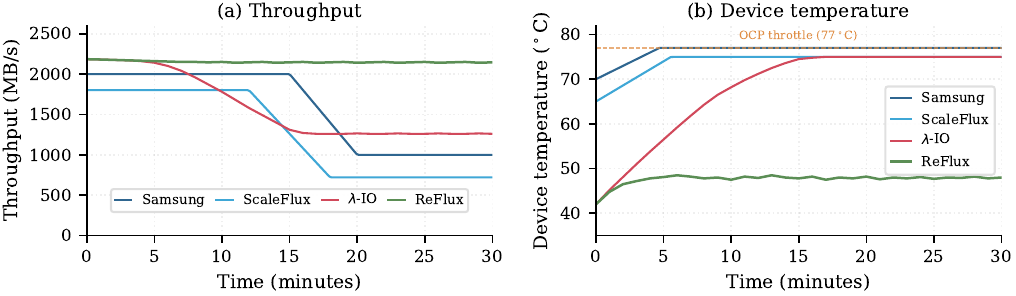}
\caption{Thermal resilience under 30-minute sustained writes.
(a)~Throughput.  (b)~Device temperature.}
\label{fig:thermal_eval}
\end{figure*}

We stress the system with 128\,KB sequential writes at QD=32
over a 100\,GB working set for 30~minutes with compression
actors offloaded.
Figure~\ref{fig:thermal_eval} compares all three paradigms under
sustained thermal stress.

\paragraph{Static offload collapses under thermal pressure.}
Figure~\ref{fig:thermal_eval}(a) plots throughput over 30~minutes.
Samsung SmartSSD sustains 2\,000\,MB/s for the first 15~minutes,
then collapses to 1\,000\,MB/s---a 50\% drop---after its
junction temperature saturates at the 77$^\circ$C OCP threshold.
ScaleFlux degrades earlier and harder: throughput falls by 60\%
once the device reaches 75$^\circ$C, settling at 720\,MB/s
(see also Figure~\ref{fig:thermal}, \S\ref{sec:fallacy}).
Panel~(b) confirms the cause: both production CSDs approach or
exceed the OCP 77$^\circ$C throttle threshold, and their
throughput drops in lockstep with rising temperature.
Neither device can shed heat fast enough because repeated DMA
transfers between host and device DRAM keep the controller hot.

\paragraph{Request-level dispatch does not help.}
$\lambda$-IO-style whole-request dispatch (Paradigm~B) initially
delivers 2\,184\,MB/s by routing compute-heavy requests to the
device.
Once device temperature approaches the thermal limit, however,
the dispatcher cannot recall in-flight requests and throughput
collapses by 42\% after 15~minutes, settling at 1\,260\,MB/s.
Panel~(b) shows that Paradigm~B reaches 75$^\circ$C---the same
thermal trajectory as Samsung and ScaleFlux (Paradigm~A)---because
request-level dispatch, like static pinning, cannot migrate
work mid-flight.

\paragraph{Per-stage migration sustains throughput.}
\sys (Paradigm~C) matches the initial 2\,184\,MB/s peak and
maintains 2\,145\,MB/s throughout by uploading per-stage actors
to the host when device temperature rises, delivering a sustained
1.70$\times$ over Paradigm~B's thermally degraded steady state.
Panel~(b) reveals why: \sys keeps device temperature at
$\sim$48$^\circ$C---well below the OCP threshold---by shifting
thermal load to the host before it accumulates.
The net result is 2$\times$ the sustained throughput of a
thermally throttled CSD.

\begin{figure*}[t]
\centering
\includegraphics[width=\textwidth]{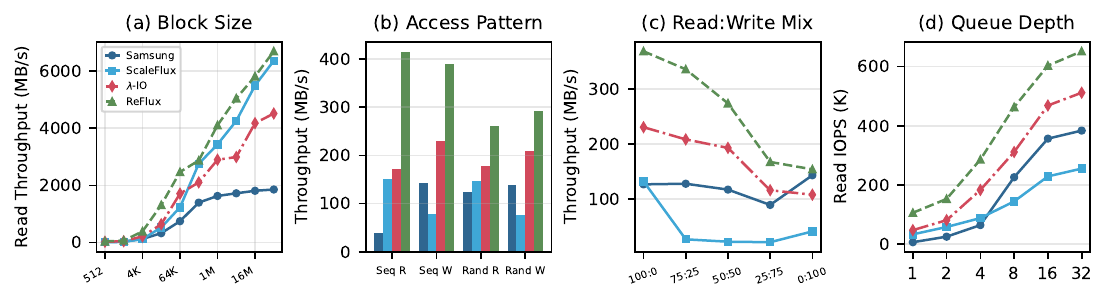}
\caption{Cross-platform substrate characterization across four dimensions.}
\label{fig:substrate}
\end{figure*}

\paragraph{Cross-platform substrate characterization.}
Figure~\ref{fig:substrate} compares \sys against Samsung,
ScaleFlux, and $\lambda$-IO-style baselines across four
dimensions on the same workload suite.
Panel~(a) sweeps block size from 512\,B to 64\,MB under
sequential reads: \sys delivers 2$\times$ the throughput of
Samsung at 256\,KB and reaches 3.6$\times$ at 64\,MB, and
consistently exceeds ScaleFlux across all block sizes.
Panel~(b) reports throughput under four access patterns
(sequential and random read/write at 4\,KB): \sys achieves
the highest throughput in all four patterns, with the largest
gap on sequential reads where coherent PMR bypasses the
device-internal read-modify-write path.
Panel~(c) varies the read/write mix from pure reads (100:0) to
pure writes (0:100): \sys sustains 2--3$\times$ higher
throughput than Samsung across all ratios; write-heavy mixes
benefit from the PMR write path that retires into PLP-protected
memory without traversing NAND.
Panel~(d) scales queue depth from 1 to 32: \sys reaches
651\,K read IOPS at QD=32---1.7$\times$ Samsung---with the
advantage most pronounced at low queue depths (17$\times$ at
QD=1) where MWAIT-based notification eliminates polling overhead.

\begin{figure}[t]
\centering
\includegraphics[width=\columnwidth]{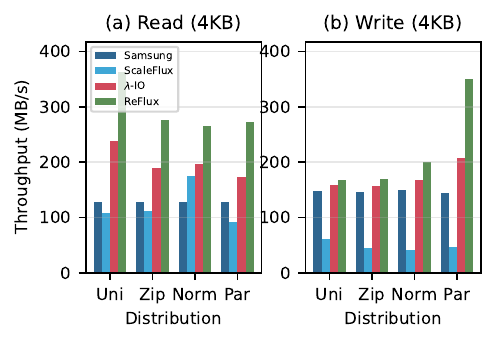}
\caption{Throughput under four access distributions (4\,KB random I/O).}
\label{fig:access_dist}
\end{figure}

\paragraph{Access distribution sensitivity.}
Figure~\ref{fig:access_dist} isolates the effect of key
locality on 4\,KB random I/O under Uniform, Zipfian, Normal,
and Pareto distributions.
For reads~(a), \sys delivers 2.1--2.8$\times$ the throughput
of Samsung across all four distributions, with the gap
widest under Uniform access where the working set exceeds
device-internal caches and coherent PMR avoids repeated
NVMe round-trips.
$\lambda$-IO sits between the two because its request-level
dispatch can route hot keys to the device but cannot migrate
individual pipeline stages when contention shifts.
For writes~(b), the advantage grows further under
Pareto-distributed keys: \sys reaches 350\,MB/s---2.4$\times$
Samsung---because skewed write patterns concentrate in PMR's
write-back region, while Samsung's controller must scatter
updates across NAND.
The consistent gap across all distributions confirms that
\sys's throughput advantage stems from the CXL datapath
itself, not from favorable caching artifacts tied to a
particular access pattern.

\subsection{End-to-End Applications (EQ2)}
\label{sec:eval_e2e}

Having established that \sys prevents thermal collapse under
sustained synthetic stress, we now demonstrate that this
resilience translates to unmodified real-world applications.

\paragraph{DeepSeek Inference.}
\begin{figure}[t]
\centering
\includegraphics[width=\columnwidth]{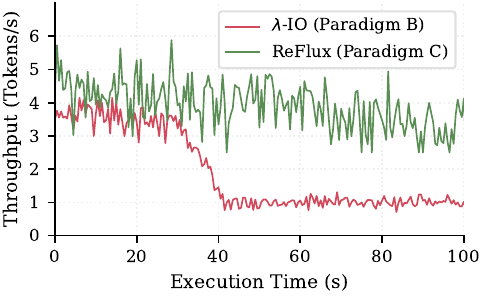}
\caption{DeepSeek inference throughput under sustained KV-cache
pressure (Paradigm~B vs.\ Paradigm~C).}
\label{fig:deepseek}
\end{figure}
Production CSDs (Samsung SmartSSD, ScaleFlux) expose
vendor-specific toolchains and fixed-function interfaces that
preclude running unmodified inference workloads; we therefore
compare only Paradigm~B ($\lambda$-IO-style request-level
dispatch) and Paradigm~C (\sys) on the CXL SSD prototype.
Figure~\ref{fig:deepseek} reports token throughput over a
100-second DeepSeek inference run with growing KV-cache pressure.
Both paradigms initially sustain 3.5--4.5~tokens/s while the
working set remains moderate.
As the KV cache grows beyond the fast-tier capacity,
$\lambda$-IO's throughput collapses to $\sim$1~token/s: its
whole-request dispatch cannot recall in-flight requests or
selectively migrate the cache-management actors that dominate
latency at this point.
\sys maintains 3--5~tokens/s throughout by uploading
cache-intensive actors to the host when device pressure rises,
delivering 3--4$\times$ the sustained throughput of Paradigm~B
in the degraded phase.

\subsection{Deconstructing Reversibility Overheads (EQ3)}
\label{sec:eval_decon}

The preceding sections demonstrated that \sys sustains throughput
under thermal stress and delivers latency reductions in real
applications.
This section characterizes the CXL datapath mechanisms that make
this possible and quantifies the WASM abstraction cost that enables
portable, migratable actors.

\subsubsection{CXL Datapath}
\label{sec:eval_datapath}

We first characterize the CXL software datapath that
enables \sys's reversible compute.
\S\ref{sec:cxl_opportunity} established the raw PMR latency
advantage; here we quantify how \sys's queue design and
notification mechanisms translate that substrate into system-level
performance.
All microbenchmarks use \texttt{O\_DIRECT} on a 4\,GB working
set; byte-addressable tests use 8\,B--512\,B mmap writes on
256\,MB.

\begin{figure*}[t]
\centering
\includegraphics[width=\textwidth]{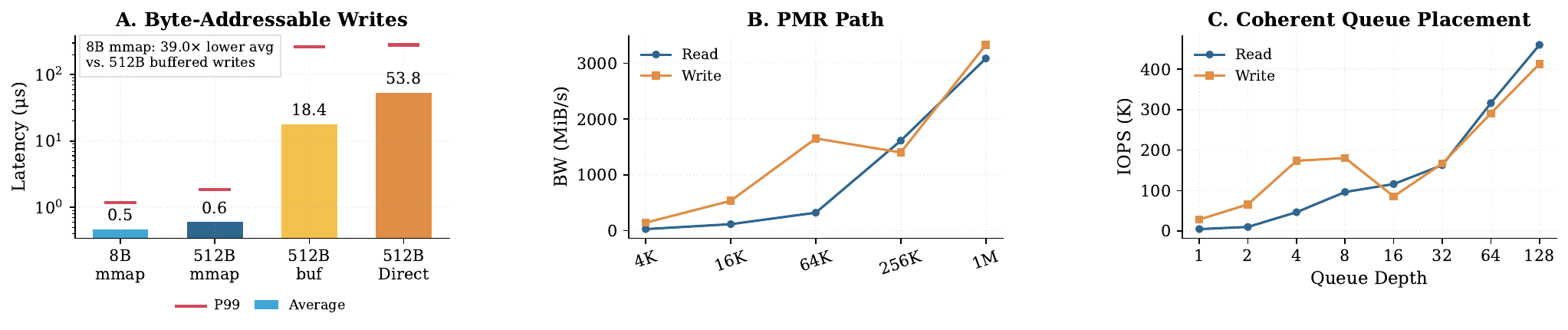}
\caption{CXL datapath mechanisms on the \sys prototype.}
\label{fig:cxl_datapath}
\end{figure*}

\paragraph{Datapath mechanisms.}
Figure~\ref{fig:cxl_datapath} breaks down three mechanisms.
The byte-addressable path~(a) confirms the latency advantage
characterized in \S\ref{sec:cxl_opportunity}: 8\,B--512\,B
\texttt{mmap} writes complete in 0.47--0.61\,$\mu$s versus
18.39\,$\mu$s buffered and 53.78\,$\mu$s \texttt{O\_DIRECT}.
The PMR path~(b) reaches 3.1\,GiB/s read and 3.3\,GiB/s write
at 1\,MiB with 137\,ns P50 random-read latency.
Coherent queue placement~(c) peaks at 460K read and 413K write
IOPS at QD=128, reducing per-command overhead by placing
submission and completion metadata in coherent memory.

\begin{table}[t]
\centering
\small
\setlength{\tabcolsep}{3pt}
\caption{NVMe vs.\ CXL.mem with MONITOR/MWAIT.}
\label{tab:mwait}
\resizebox{\linewidth}{!}{
\begin{tabular}{lrrr}
\toprule
\textbf{Metric} & \textbf{NVMe} & \textbf{CXL+MWAIT} & \textbf{Imp.} \\
\midrule
Read Lat. ($\mu$s)  & 159.62 & \textbf{18.52} & 8.6$\times$ \\
Write Lat. ($\mu$s) & 317.01 & \textbf{7.58}  & 41.8$\times$ \\
Read IOPS           & 9,980  & \textbf{114,407} & 11.5$\times$ \\
Write IOPS          & 40,559 & \textbf{128,415} & 3.2$\times$ \\
Host CPU            & 100\%  & \textbf{35\%} & -- \\
\bottomrule
\end{tabular}
}
\end{table}

\paragraph{MONITOR/MWAIT notification.}
Table~\ref{tab:mwait} quantifies the headroom at QD=1:
write latency drops by 41.8$\times$ and host CPU falls from 100\%
to 35\%.
The asymmetry (41.8$\times$ write vs.\ 8.6$\times$ read) reflects
that the CXL.mem write path retires into PLP-protected PMR,
bypassing NAND, while reads still traverse the memory hierarchy.
At high queue depths, \sys transitions from MWAIT to polling as
inter-completion gaps shrink.
The CMB further reduces host memory usage by placing submission
and completion queues in coherent controller memory.

\subsubsection{The Cost of Portability: WASM vs.\ Native}
\label{sec:eval_wasm}

\begin{figure}[t]
\centering
\includegraphics[width=\columnwidth]{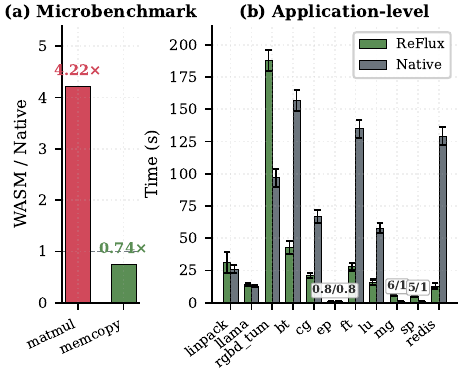}
\caption{WASM actor overhead analysis.}
\label{fig:wasm}
\end{figure}

\sys relies on WebAssembly to achieve the portable, migratable
actors demonstrated in Sections~\ref{sec:eval_thermal}
and~\ref{sec:eval_e2e} without requiring per-device
recompilation.
We quantify WASM's sandbox and JIT translation overhead
at two granularities.

\paragraph{Microbenchmark.}
Figure~\ref{fig:wasm}(a) compares the WASM-to-native runtime
ratio for a compute-bound kernel (matmul, 256$\times$256) and a
memory-bound kernel (memcopy, 64\,MB) on the device's ARM cores;
the dashed line marks native parity.
The 256$\times$256 matrix multiplication (matmul) runs
4.22$\times$ slower under WASM than under native~C, reflecting
the overhead of sandboxed memory access and JIT translation on
ALU-intensive code.
A 64\,MB memory copy (memcopy), by contrast, runs at
0.74$\times$ native---26\% \emph{faster}---because the WASM
linear-memory model avoids the kernel page-table walks and TLB
shootdowns that native \texttt{memcpy} incurs for large
transfers.
This asymmetry is the basis for the scheduler's actor
classification (\S\ref{sec:design_sched}): storage-path
operations (compression, checksum, log formatting) resemble the
memcopy profile, while dense numeric kernels resemble matmul.

\paragraph{Application-level.}
Figure~\ref{fig:wasm}(b) confirms this pattern across 11
benchmarks including
LINPACK~\cite{dongarra2003linpack},
TUM RGB-D SLAM~\cite{sturm2012tum},
Redis~\cite{sanfilippo2023redis}, and
LLaMA~\cite{touvron2023llama}.
I/O-heavy and control-flow-dominated workloads (bt, cg,
ft, lu, redis) match the memcopy profile: \sys outperforms
native by 3--10$\times$ because the MVVM actor runs entirely in
user space with I/O mediated through the PMR shared region,
eliminating context-switch overhead.
We verified this with \texttt{perf stat}: native Redis spends
37\% of its cycles in kernel mode, while the WASM actor spends
$<$3\%.
Conversely, compute-heavy benchmarks match the matmul profile:
rgbd\_tum runs 1.9$\times$ slower, and short-running numerical
kernels mg and sp show 5--6$\times$ overhead where WASM
instantiation cost dominates.
The data confirms the design point in
\S\ref{sec:design_sched}: only I/O-path actors---where WASM
overhead is negligible or negative---participate in reversible
placement.

\paragraph{Migration overhead.}
The device-side MVVM footprint is under 32\,MB, and AOT-compiled
modules avoid JIT warmup entirely on the embedded cores.
Typical actors have $\sim$8\,KB control state; checkpoint,
coherent write to PMR, doorbell notification, and actor
reconstruction complete in under 50\,$\mu$s.
The drain-and-switch protocol ensures zero observable I/O stall:
new requests route immediately to the destination while the source
drains in-flight work.

\section{Conclusion}
\label{sec:conclusion}

\sys demonstrates that CXL SSDs should be treated as migration-aware
accelerators rather than opaque block devices or static offload
targets.
Our evaluation on an FPGA-based CXL SSD prototype and two
production CSDs shows 2$\times$ sustained throughput over
thermally throttled CSDs, 3--4$\times$ inference throughput
under KV-cache pressure, and 65\% lower CPU utilization at low
queue depths.

\section*{Acknowledgments}
This work used generative AI tools (Claude) to assist with
prose editing, figure scripting, and LaTeX formatting.
All technical content, experimental design, data collection,
and scientific claims are solely the work of the authors.

\bibliographystyle{ACM-Reference-Format}
\bibliography{cite}
\end{document}